\documentclass[epjCONF]{svjour}
\usepackage{graphics}
\usepackage[varg]{txfonts} 
\usepackage[latin1]{inputenc}
\session-title{MESON2012 - 12th International Workshop on Meson Production, Properties and Interaction} 
\begin{document}
\title{Strangeness production in antiproton-nucleus annihilation}
\author{A.B. Larionov\inst{1,2,3}\fnmsep\thanks{\email{Alexei.Larionov@theo.physik.uni-giessen.de}}
   \and T. Gaitanos \inst{1} \and H. Lenske \inst{1} \and U. Mosel \inst{1}}
\institute{Institut f\"ur Theoretische Physik, Universit\"at Giessen,
             D-35392 Giessen, Germany \and National Research Center ``Kurchatov Institute'', 
             123182 Moscow, Russia \and Frankfurt Institute for Advanced Studies, J.W. Goethe-Universit\"at,
             D-60438 Frankfurt am Main, Germany  }
\abstract{
The results of the microscopic transport calculations of 
$\bar p$-nucleus interactions within a GiBUU model are presented. 
The dominating mechanism of 
hyperon production is the strangeness exchange processes 
$\bar K N \to Y \pi$ and $\bar K N \to \Xi K$.
The calculated rapidity spectra of $\Xi$ hyperons are significantly 
shifted to forward rapidities with respect to the spectra of $S=-1$ 
hyperons.
We argue that this shift should be a sensitive test for the possible 
exotic mechanisms of $\bar p$-nucleus annihilation. The production
of the double $\Lambda$-hypernuclei by $\Xi^-$ interaction with a 
secondary target is calculated. 
}

\maketitle

\section{Introduction}

\label{intro}

The interest to strangeness production in $\bar p$-nucleus interactions
was originally related to the mechanism of strangeness enhancement 
in a quark-gluon plasma (QGP) proposed by Rafelski and M\"uller in early 
80's for relativistic heavy-ion collisions \cite{Rafelski82}.
This idea has driven several experiments at BNL \cite{Condo84,Ahmad97}, LEAR 
\cite{Balestra87} and KEK \cite{Miyano88}. 
Although the following-up theoretical analyses within the 
intranuclear cascade (INC) models \cite{Cugnon90,Gibbs90} seem to support the usual mechanism of 
strangeness production in terms of binary hadron-hadron collisions, the collected experimental
data constitute a very useful base for testing newly developing theoretical models needed
in view of forthcoming experiments with antiproton beams at FAIR.
In this talk we discuss our recent results of the microscopic 
transport calculations of the $K^0_S$, $\Lambda$- and $\Xi^-$-hyperon and
double-$\Lambda$ hypernuclei production based on the Giessen 
Boltzmann-Uehling-Uhlenbeck (GiBUU) transport model. 
Sec. 2 contains some model details. Sec. 3 collects the results of our 
calculations. We summarize in sec. 4.     

\section{Model}

\label{model}

The GiBUU model \cite{GiBUU} is a unified transport-theoretical approach capable
to describe photon-, electron-, neutrino-, hadron- and nucleus-induced reactions
on nuclei. To formulate transport equations, we are using here a relativistic 
mean-field model.
The transport equations for the different 
baryons $i=N,~N^*,~\Delta,~Y,~\Xi,$..., respective antibaryons and mesons
$~\pi,~\eta,~\rho,~\omega,~K,~\bar K,$... can be written
as follows: 
\begin{equation}
  (p_0^*)^{-1}
  \left[ p^{*\mu} \partial^x_\mu + \left(p_\nu^* F_i^{\mu\nu} 
                                   + m_i^* \partial_x^\mu m_i^*\right)
    \partial^{p^*}_\mu \right] f_i(x,{\bf p^*}) 
  = I_i[\{f\}]~.                                                 \label{kinEq}
\end{equation}
Here, $f_i(x,{\bf p^*})$ is the phase-space distribution (or Wigner) function, $p^*=p-V_i$ is 
the kinetic four-momentum, $F_i^{\mu\nu} \equiv \partial^\mu V_i^\nu - \partial^\nu V_i^\mu$ 
is the field tensor, and $m_i^*=m_i+S_i$ is the effective mass. The baryonic mean field
is characterized by the scalar, $S_i$, and vector, $V_i^\mu$, potentials.
The collision term in the r.h.s. of Eq.(\ref{kinEq}) describes the residual two-body
interactions beyond mean field and resonance decays. The two-body collision
term depends on the angular differential cross sections of the particle-particle
scattering and takes into account the Pauli blocking factors for the outgoing nucleons.
The following relevant for the present study collision channels are included:
annihilation $\bar N N \to$ mesons, $\bar N N \to \bar N N$, 
$\bar N N \to \bar N \Delta$ (+c.c.), 
$\bar N N \to \bar\Lambda \Lambda,~\bar\Sigma \Lambda (+{\rm c.c}),~\bar\Xi \Xi$.
Hyperons are also produced in strangeness exchange reactions on nucleons
$\bar K N \to Y \pi$, $\bar K N \to \Xi K$, and in the collisions of nonstrange
mesons with nucleons $M N \to Y K$, $M=\pi,~\eta,~\rho,~\omega$. The produced hyperons 
may rescatter and change their charge and/or flavour via the following processes: 
$\Lambda N \to \Lambda N$, $\Lambda N \leftrightarrow \Sigma N$, $\Sigma N \to \Sigma N$, 
$\Xi N \to \Xi N$, $\Xi N \to \Lambda \Lambda$, $\Xi N \to \Lambda \Sigma$.
Further details of the model can be found in \cite{GiBUU,Larionov12,Gaitanos12}.

\section{Results}

\label{results}

\begin{figure}
\begin{center}
   \resizebox{0.75\columnwidth}{!}{\includegraphics{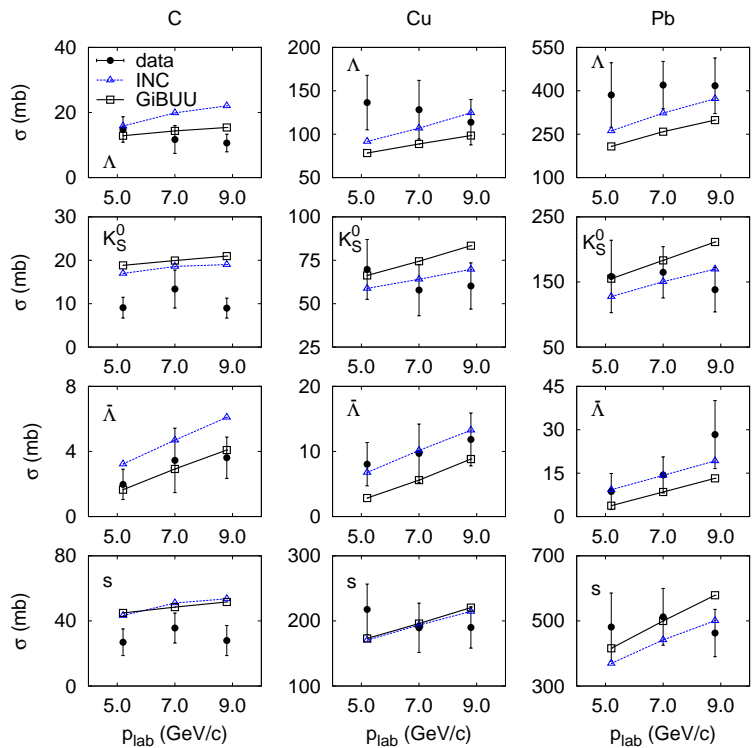}}
\end{center}
\caption{Beam momentum dependence of $\Lambda$, $K_S^0$, $\bar\Lambda$ and strange 
quark production cross sections in $\bar p$ interactions with $^{12}$C, $^{64}$Cu 
and $^{208}$Pb nuclei. The calculated $\Lambda$ production cross sections include
also the contribution of $\Lambda$'s from $\Sigma^0 \to \Lambda \gamma$ decays.
Figure taken from \cite{Larionov12}.}
\label{fig:sig_str_vs_plab}
\end{figure}
Fig.~\ref{fig:sig_str_vs_plab} shows our results on the inclusive cross sections of
neutral strange particle production in comparison with experimental data 
\cite{Ahmad97} and INC model calculations \cite{Gibbs90}. There is an overall
satisfactory agreement of GiBUU calculations with data and with INC results.
We observe, however,
a systematic trend to underestimate $\Lambda$ production for heavier targets and
overestimate $K_S^0$ production for light targets by transport calculations.
The reason for this is not yet clear for us. 
One possibility is that $\bar K$ absorption cross sections in nuclear medium
are enhanced. This suggestion is based on our observation that 
$\sim 60-80$\% of the $S=-1$ hyperon production rate is due to $\bar K N \to Y X$,
$\bar K N \to Y^*$ and $\bar K N \to Y^* \pi$ reactions.

\begin{figure}
\begin{center}
   \resizebox{0.75\columnwidth}{!}{\includegraphics{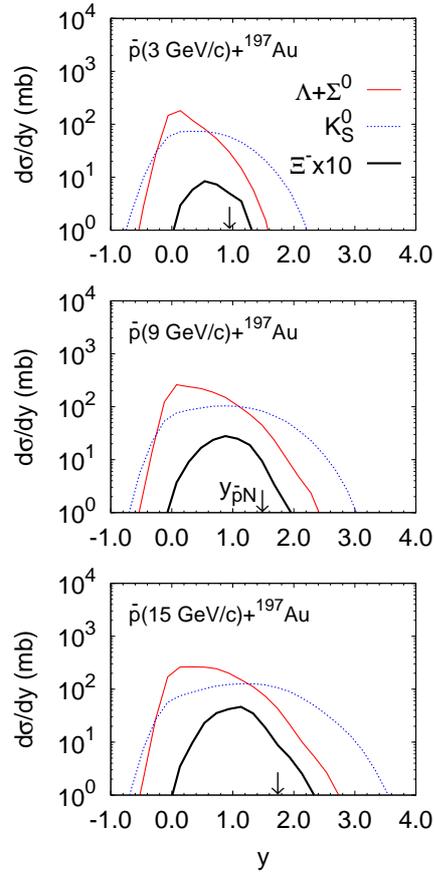}}
\end{center}

\vspace*{-1cm}

\caption{Rapidity distributions of $\Xi^-$, $\Lambda$ and $K^0_S$ in
$\bar p+^{197}$Au collisions at 3, 9 and 15 GeV/c. The $\bar p N$ center-of-mass
(c.m.) rapidities are marked with arrows. Figure taken from \cite{Larionov12}.}
\label{fig:dsig_dy_pbarAu}
\end{figure}
In Fig.~\ref{fig:dsig_dy_pbarAu} we present the rapidity spectra of 
$\Xi^-$-hyperons plotted together with the rapidity spectra of $\Lambda$-hyperons 
and $K_S^0$'s.
The $\Lambda$ rapidity distributions are peaked at $y\simeq0$, because 
the $S=-1$ hyperons are mostly produced in the exothermic strangeness exchange reactions 
$\bar K N \to Y \pi$ with slow $\bar K$'s. The rescattering
on nucleons further decelerates the produced hyperons. On the other hand, 
the peaks of the $\Xi^-$ rapidity distributions are shifted forwards by 0.5-1 units
of rapidity despite that the $\Xi N$ rescattering is also included in our
calculations. This shift is easily explained by the dominating endothermic 
production channel $\bar K N \to \Xi K$ with the threshold $\bar K$ beam 
momentum of 1.048 GeV/c corresponding to the $\bar K N$ c.m. rapidity of 0.55.
We think that the difference between the peak positions of the $\Lambda$ and 
$\Xi$ hyperon rapidity spectra is the direct consequence of the underlying 
hadronic production mechanism implemented in our model and should vanish in 
the case of the strangeness production from the blob of a supecooled QGP 
\cite{Rafelski88}.

\begin{figure}
\begin{center}
   \resizebox{0.6\columnwidth}{!}{\includegraphics{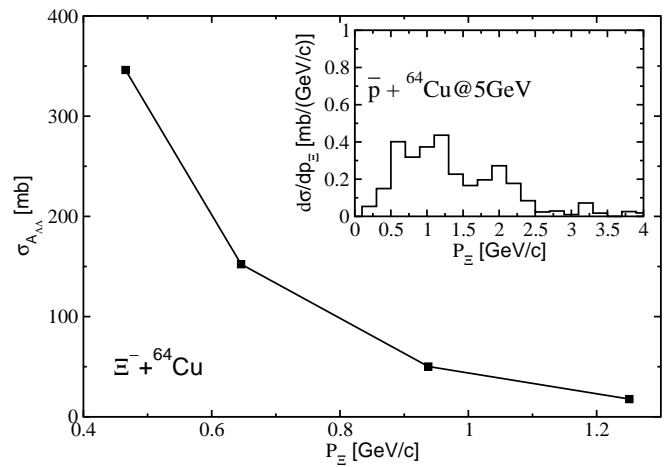}}
\end{center}
\caption{Double-$\Lambda$ hyperfragment production cross section in $\Xi^- +^{64}$Cu
interactions vs beam momentum of $\Xi^-$. For orientation, the inset shows momentum
distribution of produced $\Xi^-$'s in $\bar p+^{64}$Cu interactions at $E_{\rm lab}=5$
GeV. Figure taken from \cite{Gaitanos12}.}
\label{fig:LLprodtotal_new2}
\end{figure}

The program of the future PANDA experiment \cite{Pochodzalla05} is indended to use 
a primary target to produce $\Xi^-$ hyperons which will be then decelerated 
in the ordinary medium and captured to the Coulomb orbit of a secondary target nucleus.
The double-$\Lambda$ hypernuclear system will be created due to reaction 
$\Xi^- p \to \Lambda \Lambda$ on a proton from the secondary target nucleus. 
To get some feeling of this idea, we have performed
a simplified study by, first, calculating the momentum spectrum of emitted $\Xi^-$'s
in the primary reaction and, second, by calculating the production cross section
of the double-$\Lambda$ hyperfragments in the interaction of a $\Xi^-$ with the
secondary target. Fig.~\ref{fig:LLprodtotal_new2} shows the results. We see that
the production cross section of double-$\Lambda$ hyperfragments grows with
decreasing $\Xi^-$ beam momentum, as expected. On the other hand, as one can see
in the inset, a rather significant number of $\Xi^-$'s is produced at low
momenta ($< 1$ GeV/c). Alghough we have neglected the $\Xi^-$ deceleration in the
ordinary medium and its capture to the Coulomb orbit, our results support the
main idea of the double-$\Lambda$ cluster production experiment.
 
\section{Summary}
\label{summary}

The main conclusions of our present study are: 

\begin{itemize}

\item Overall, GiBUU is working reasonably well for strangeness 
observables in $\bar p$-induced reactions. However, $\Lambda$ 
production is slightly underpredicted and $K^0_S$ production 
is overpredicted. 

\item The $S=-2$-hyperon rapidity spectra are sensitive to the
underlying production mechanism: by hadronic two-body collisions
or by QGP hadronization.

\item Double-$\Lambda$ hyperfragment production cross section due to 
$\Xi^-$ interaction with a secondary target grows with decreasing
$\Xi^-$ beam momentum reaching hundreds mb below 0.8 GeV/c.

\end{itemize}

This work has been financially supported by BMBF, HIC for FAIR and DFG (Germany), 
and Grant NSH-7235.2010.2 (Russia).

\end{document}